\documentclass[sigconf,nonacm]{acmart}

\usepackage{verbatim}
\usepackage{subcaption}
\usepackage{todonotes}
\usepackage{amsfonts}
\usepackage{bbm}
\usepackage{csquotes}
\usepackage[inline]{enumitem}

\presetkeys%
    {todonotes}%
    {inline,backgroundcolor=yellow}{}

\AtBeginDocument{%
  \providecommand\BibTeX{{%
    \normalfont B\kern-0.5em{\scshape i\kern-0.25em b}\kern-0.8em\TeX}}}

\setcopyright{acmcopyright}
\copyrightyear{2021}
\acmYear{2021}
\acmDOI{}

\usepackage{xpatch}

\setcopyright{none}
\acmConference[SimuRec '21]{SimuRec '21: Workshop on Simulation Methods for Recommender Systems at ACM RecSys}{September 27th - October 1st, 2021}{Amsterdam, NL}
\acmBooktitle{SimuRec '21: Workshop on Simulation Methods for Recommender Systems at ACM RecSys,
  September 27th - October 1st, 2021, Amsterdam, NL}
\acmPrice{15.00}
\acmISBN{978-1-4503-XXXX-X/18/06}

\begin{document}

\title[Understanding Longitudinal Dynamics of Recommender Systems]{Understanding Longitudinal Dynamics of Recommender Systems with Agent-Based Modeling and Simulation}\titlenote{Paper accepted for the Workshop on Simulation Methods for Recommender Systems (SimuRec) at ACM RecSys '21.}

\author{Gediminas Adomavicius}
\affiliation{%
  \institution{University of Minnesota}
}
\authornote{Authors are listed in alphabetical order.}
\email{gedas@umn.edu}

\author{Dietmar Jannach}
\affiliation{%
  \institution{University of Klagenfurt}
}
\email{dietmar.jannach@aau.at}

\author{Stephan Leitner}
\affiliation{%
  \institution{University of Klagenfurt}
}
\email{stephan.leitner@aau.at}

\author{Jingjing Zhang}
\affiliation{%
  \institution{Indiana University}
}
\email{jjzhang@indiana.edu}

\renewcommand{\shortauthors}{Adomavicius et al.}

\begin{abstract}
Today's research in recommender systems is largely based on experimental designs that are static in a sense that they do not consider potential longitudinal effects of providing recommendations to users. In reality, however, various important and interesting phenomena only emerge or become visible over time, e.g., when a recommender system continuously reinforces the popularity of already successful artists on a music streaming site or when recommendations that aim at profit maximization lead to a loss of consumer trust in the long run. In this paper, we discuss how \emph{Agent-Based Modeling and Simulation (ABM)} techniques can be used to study such important longitudinal dynamics of recommender systems. To that purpose, we  provide an overview of the ABM principles,  outline a simulation framework for recommender systems based on the literature, and discuss various practical research questions that can be addressed with such an ABM-based simulation framework.
\end{abstract}


\maketitle

\section{Introduction and Motivation}
Recommender systems can exert significant influence on how users navigate information spaces and make decisions. In parallel, these changes in user behavior impact other stakeholders as well, such as the providers of recommendation services who, for example, seek to obtain higher customer retention rates or more sales through the recommendations \cite{abdollahpouri2020,jannachjugovactmis2019}. In practice, providers of recommendation services monitor the effects of the introduction or modification of a recommender system over an extended period of time, often within A/B tests (i.e., field experiments), because only longer observation periods may lead to reliable insights regarding the sustainability of the desired effects.  However, field experiments have their limitations; e.g., in real-world experimental settings it may be difficult (or even impossible) to control for the effects of various confounding factors, such as users' self-selected consumption strategies \cite{consumptionPerformance2020}.

In academia, where access to major deployed systems is often not available, various proxies for assessing the effects of different recommendation algorithms or system-user interfaces can be used. The predominant research method is \emph{offline} experiments, where we use datasets containing user preference information, withhold some of the data, and then rely on various computational metrics, e.g., to assess the prediction performance of an algorithm or its capability to generate diversified recommendations. Such analyses are without doubt useful to compare algorithms in certain respects. However, a major limitation is that in experiments of this type we can only generate and evaluate recommendations for each user based on a static set of data.
As such, these experiments do not allow us to investigate the potentially undesired longitudinal effects of personalized recommendations in practice. An example of such an effect is when a recommender systems reinforces the popularity of already popular items over time, ultimately leading to \emph{decreased} sales diversity \cite{LeeHosanagar2018,JannachLercheEtAl2015,DBLP:conf/icis/LeeH14}.
Controlled studies with users are an alternative to offline experiments. Such studies commonly focus on user perceptions of the quality of system-generated recommendations, see, e.g., \cite{Ekstrand:2014:UPD:2645710.2645737}. Again, however, user studies typically consist of a relatively short (often one-time) interaction of the participants with the system. While sometimes participants are queried about their intention to use the system in the future \cite{Pu2011}, longitudinal effects are typically not examined.

Regardless of the applied methodology, most research today focuses on evaluating recommender systems from the consumer side, e.g., in terms of reduced search effort or the probability of discovering new items. In reality, however, besides consumers, recommender systems are also designed to serve the content providers and additional stakeholders \cite{abdollahpouri2020,JannachAdomavicius2016}. As a result, the recommendations may affect different stakeholders, and they may again lead to phenomena that only emerge or become visible over time. For example, an e-commerce platform might try to recommend items that are favorable in terms of the profit margin. Due to the persuasive effect of recommendations, initially consumers might indeed purchase the recommended items more often. Over time, however, consumers might lose their trust in the recommender or even the e-commerce platform as a whole due to repeated negative experiences. Again, such phenomena might not be easily analyzed using the traditional experimental approaches discussed above.

In this paper, we argue that several of these aspects can be investigated with the help of \emph{Agent-based Modeling (ABM) and Simulation} \cite{Bonabeau2002,MillerPage2007}. ABM has been successfully applied in a variety of research fields, including Social Sciences, Economics, and Information Systems, to model complex adaptive systems. It is based on simulating the actions of heterogeneous and autonomous agents, with the goal of understanding the effects of these actions and the interactions on the entire system in a longitudinal perspective. Thus, ABM is a promising approach that is emerging in recommender systems literature for investigating questions related to dynamics or evolution of a system and its impact over time.

\section{Agent-Based Modeling and Simulation}
\label{sec:ABM-background}
ABM is an approach to model complex adaptive systems that are composed of three main building blocks:
\begin{enumerate*}[label=\textit{(\roman*)}]
\item heterogeneous and autonomous \enquote*{agents},
\item the environment in which agents operate, and
\item interactions among agents and among agents and the environment \cite{wall2020,Macal2010,Leitner2015}.
\end{enumerate*}
ABM follows a \textit{generative science} approach \cite{Epstein2006}: The modeling process takes place at micro-level, i.e., at the level of the three building blocks, and the ultimate goal of the simulation typically is to obtain macro-level insights from micro-level interactions \textit{over a longer period of time}, i.e., insights about the collective behavior and emerging phenomena.

The dynamics within agent-based models (ABMs) are mainly driven by the set of %
\emph{agents} who are usually heterogeneous and autonomously make decisions and take actions. Agents represent either real-world or artificial entities. Real-world agents 
in our case could include providers of recommendation services, users of such services, or organizations that provide the items that are subject to recommendation. These real-world agents are described in terms of specific characteristics, such as objectives, rules to autonomously make decisions, emotional states, and capabilities to collect, process, and interpret information \cite{Tesfatsion2006,wall2020}. Artificial agents, on the other hand, could represent bots or autonomous decision-making entities that are fed by the methods of AI and machine learning.
Agents operate in
an \emph{environment}. What exactly is represented by the environment strongly depends on the research objective and can, for example, be the social context in which agents interact, an actual physical space in which users move around, or the technical environment in which recommendations are provided and received \cite{Sassi2017}.
\emph{Interactions} can either take place directly by specified communication rules or indirectly via the environment. Direct communication could cover the exchange of information between agents using specified (direct) communication channels. In a recommendation service context, the directly exchanged information can include users' revealed preferences, perceived usefulness of recommendations, or the trust in the service provider, while indirect communication could be represented by the public opinion that is formed in social networks and emerges from opinion dynamics \cite{Acemoglu2011}. I.e., the researcher has to specify the \enquote*{topology of interactions} which, aside from the structure itself, also includes to model the temporal sequence of interactions \cite{wall2020}.

It is important to note that ABM systems can be adaptive, i.e., the population of agents and their characteristics, the environment, and the topology of interactions might change over time \cite{Holland1992}. In the context of recommendation services, the adaptation might take the form of consumers who learn information and revise their beliefs or adapt their consumption behavior, recommendation service providers that might develop new recommendation strategies, and environments that change as new competitors enter the market or new social networks form. Also the topology of interactions might be subject to change, e.g., as new communication channels emerge or the agents autonomously decide to interact more or less frequently. Once the building blocks of ABM are defined, the model is implemented in software and \enquote*{solved} via simulation, so that the emerging patterns can be derived from extensive numerical results~\cite{chang2006}.

ABM allows for rich contextualization and for exploring and isolating the effects of different factors \cite{wall2020,consumptionPerformance2020}. Modelling at the micro-level offers exceptional control which not only enables researchers to model complex individual behaviors  (e.g., to integrate rich assumptions about the users and interactions in terms of boun-ded rationality and cognitive biases, and network topology, respectively), but also allows these behaviors of agents to be isolated and specified with precision. Thus, ABM appears to be particularly suitable if \textit{realistic} real-world agents should be captured. Studying the emergence of longitudinal effects could, theoretically, also be achieved with longitudinal field experiments; however, they might be prohibitively expensive. In contrast, ABM provides a viable and scalable methodology to analyze longitudinal effects in a recommendation context \cite{consumptionPerformance2020}.

\section{Studying Dynamics of Recommender Systems with Agent-based Modeling}
\label{sec:ABM-in-RS}

Generally, simulation-based methodology has been used for studying various research questions in the recommender systems literature, including the effect of recommendations on product sales diversity \cite{FlederBlockbuster2009}, the robustness of recommendation strategies to artificial manipulations \cite{prawesh2014most}, the algorithmic confounding in recommender systems \cite{chaney2018algorithmic}, and the recommender effects on news consumption \cite{bountouridis2019siren}.  In addition, several studies have specifically focused on ABM-based simulation methodology, leading to insights about longitudinal dynamics of recommender systems \cite{adomavicius2013understanding,consumptionPerformance2020,longitudinalimpact2021}. In particular,  \cite{adomavicius2013understanding,consumptionPerformance2020} show that, if users overly rely on the recommendations provided by a system for their consumption choices, then, paradoxically, the system's quality will diminish over time (e.g., as measured by the system's prediction accuracy and users' consumption diversity). Further analyses reveal that a hybrid strategy that combines personalized and (popular) non-personalized item suggestions can lead to a sustainable relevance of the recommendations.  The work presented in \cite{longitudinalimpact2021} further builds on the same ABM approach, but focuses on the longitudinal effects of \emph{preference biases}. Such effects may occur when the feedback ratings that consumers give on items do not represent their true preferences but are biased, for example, by the system's predicted ratings that are shown to consumers along with the suggested items. Among other aspects, the simulations revealed that such \enquote*{polluted} preferences are subsequently reflected in the system's future predictions, leading to a feedback loop where continuously more noisy data is propagated in the system.

The works in \cite{consumptionPerformance2020,longitudinalimpact2021} are based on a general-purpose ABM framework for recommender systems modeling and simulation. The main components of the framework are the users (consumers), the recommendable items, and the recommender engine. The framework allows to model the dynamics of the \emph{user agents} using several key aspects of user behaviors as related to recommender systems: the lifespan when users are active in the system, users' preferences for items, users' item consumption frequencies, their consumption choice strategies, their rationale when providing feedback or the likelihood that they will provide feedback. The properties and dynamics of the item catalog include the point in time when items become available, their lifespan, and their \emph{content} descriptions. The modeling of user and item aspects allow the framework to properly and robustly generate `ground truth' user preferences for any consumed item. Finally, modeling the \emph{recommender engine}, also an agent in the system, involves describing both its functionality (e.g., in terms of its relevance prediction, item ranking, and performance assessment methods) and a representation of the recommender system's state at a certain point in time (e.g., in terms of the current rating database).

Given the individual framework components of \cite{consumptionPerformance2020,longitudinalimpact2021}, an iterative simulation procedure is applied. First, the system is initialized with the user population, the item catalog, the rating database and an initial set of recommendations for each user. Additional elements could be added to this environment, for example, the social network of the users (with its corresponding social influences). In the main simulation loop, a subset of the users consume one of the recommended items with a certain probability (based on the desired choice strategy); after item consumption, the user determines her preference rating for this item and, based on some probability, decides to submit her feedback to the recommender engine. This represents one form of interaction in this framework. Based on this feedback, the engine then updates the rating database and prepares recommendations for the next round. In this process, the system also updates the set of active users and items according to the modeled lifespans and consumption strategies of users and items. Finally, in each simulation step, the engine evaluates and logs its prediction performance and other relevant process metrics, so that performance developments can be tracked over time.

Going beyond longitudinal effects of consumption strategies \cite{consumptionPerformance2020} and preference biases \cite{longitudinalimpact2021}, the ABM framework and suitable extensions can be used to analyze a multitude of other aspects as well. For instance, in \cite{JannachLercheEtAl2015,ferrarojannachserra2020recsys} the authors analyze to what extent different recommendation algorithms lead to \emph{popularity reinforcement} (blockbuster) effect \citep{FlederBlockbuster2009}, where already popular items profit most from recommendations. They also study \emph{concentration effects}, which is a system's tendency to increasingly focus its recommendations on fewer items, thereby reducing their level of personalization over time. One main outcome of these studies was that the choice of the algorithm matters, and that certain families of algorithms are less prone to lead to such undesired effects.  Numerous other research questions related to longitudinal effects of recommender systems are still underexplored. E.g., there is little research on the long-term negative effects on the trust and behavior of users resulting from \enquote*{bad} recommendations, which could be caused by a malfunctioning system \citep{ExaminingCHAU2013180} or by the system not finding a suitable balance between item relevance and profit margin in a multi-stakeholder environment \cite{abdollahpouri2020}.

In addition, a recent emerging use of simulation in the recommender systems literature is to create environments against which reinforcement learning (RL) based recommendation algorithms can be properly benchmarked and compared. Example RL-based simulation platforms include RecSim \cite{ie2019recsim}, RecSim NG \cite{mladenov2021recsimng}, SOFA \cite{huang2020keeping}, RecoGym \cite{rohde2018recogym}, and PyRecGym \cite{shi2019pyrecgym}. Such platforms focus on simulating user feedback based on logged historical data and modeling sequential user behaviors in using RL. Using simulations allows researchers to optimize, evaluate, and compare RL-based recommendation strategies without having to run online experiments with real users. Although there are some similarities between these systems and the ABM approach, the RL-based simulation platforms are specifically designed for RL optimization and evaluation and, thus, typically are not ideal for modeling complex and heterogeneous user populations and studying emerging behaviors.

\section{Summary}
\label{sec:conclusions}
ABM is a simulation methodology that has been successfully applied in various domains to study phenomena that emerge in complex environments over time. In this paper, we argue that ABM represents a highly promising approach to study longitudinal effects of recommender systems on various stakeholders, i.e., an important and practically relevant topic that has been underexplored in research literature.

\bibliographystyle{abbrv}
\bibliography{references}

\end{document}